\documentclass[aps,prl,floats,floatfix,twocolumn,longbibliography,superscriptaddress]{revtex4-1}

\usepackage{color} %\textcolor{red} \color[RGB]{102,0,0}
\usepackage{latexsym}
\usepackage{amsmath} 
\usepackage{amssymb} 
\usepackage{bm}
\usepackage{wasysym}

\usepackage{graphicx}
\usepackage{epstopdf}
\graphicspath{{./}{./Figs/}}

\usepackage[colorlinks,linkcolor=blue,urlcolor=blue,bookmarks=true,pdftitle={}]{hyperref}

\newcommand{\beq}{\begin{eqnarray}}
\newcommand{\eeq}{\end{eqnarray}} 

\newcommand{\hide}[1]{}

\newcommand{\sect}[1]{{\bf #1.-- }}

\newcommand{\Cn}[1]{\begin{center} #1 \end{center}}

\newcommand{\hrefl}[2]{\href{#2}{(#1)}}
\newcommand{\rmrk}[1]{\textcolor{red}{#1}}

\begin{document}

\title{Chaos assisted many-body tunnelling}

\author{Urbashi Satpathi}
\affiliation{Department of Chemistry, Ben-Gurion University of the Negev, Beer-Sheva 84105, Israel}
\author{Sayak Ray}
\affiliation{Department of Chemistry, Ben-Gurion University of the Negev, Beer-Sheva 84105, Israel}
\affiliation{Physikalisches Institut, Rheinische Friedrich-Wilhelms-Universit\"at Bonn, Nu\ss allee 12, 53115, Bonn, Germany}
\author{Amichay Vardi} 
\affiliation{Department of Chemistry, Ben-Gurion University of the Negev, Beer-Sheva 84105, Israel}

%\date{\today}
\begin{abstract}

We study the interplay of chaos and tunnelling between two weakly-coupled Bose-Josephson junctions. The classical phase space of the composite system has a mixed structure including quasi-integrable self-trapping islands for particles and excitations, separated by a chaotic sea. We show that the many-body dynamical tunnelling gap between macroscopic Schr\"odinger cat states supported by these islands is chaos-enhanced. The many-body tunnelling rate fluctuates over several orders of magnitude with small variations of the system parameters or the particle number.  

\end{abstract}
%\date{}
\maketitle

%%%%%%%%%%%%%%%%%%%%%%%%%%%%%%%%%%%%%%%%%%%%%%%%%%%%%%%%%%%%%%%%%%%%%%%%%%%%%%%%%%%%%

%\sect{Introduction}
%
The realization of quantum effects on macroscopic scales stands at the forefront of modern physics. Most prominent in this respect are {\em macroscopic Schr\"odinger cat}  (MSC)  states \cite{Schroedinger35,Yurke86,Arndt99,Friedman00,Wineland13,Arndt14,Kovachy15,Hoff16,Johnson17}, defined as quantum superpositions of two macroscopically distinct many-body states. %The constituent states may be localized coherent states or Fock states in localized quasi-modes as in $N$-particle NOON and  Greenberger-Horne-Zeilinger (GHZ) states. 
Beyond fundamental studies of quantum mechanics,  the inherent many-body entanglement of MSC states is a useful resource in quantum information processing \cite{Cochrane99,Lund08,Ofek16} and quantum metrology \cite{Joo11}

Interacting many-body systems possessing symmetry exhibit doublets of MSC-like states in their energy spectrum.  For example, for sufficiently strong interactions, sections of the spectrum of a Bose-Einstein condensate confined to a symmetric double-well potential consist of odd and even macroscopic superposition states, supported by two self-trapping islands in their classical phase space \cite{Cirac98,Gordon99,Ruostekoski98,Huang06}. The macroscopic tunnelling time between the islands is inversely proportional to the doublet splitting that decreases exponentially with the  total particle number $N$, serving as an effective inverse Planck constant. This poor scaling is related to the notorious fragility of MSC states because environment-induced decoherence mixes them and collapses the MSC onto one of its classical branches \cite{Joos85,Zurek91}. Consequently, the observation of many-body tunnelling (MBT) in systems with more than a few particles remains a substantial challenge. 

%We note that the term `Macroscopic Quantum Tunnelling' has been used profusely to describe the effectively classical beating of {\em coherent} states wherein all particles populate a particular superposition of the localized quasi-modes. The tunnelling rate in this case is set by the one-particle gap between the orbitals and is hence much higher and easier to observe. Below, we use the term `Many-Body Tunnelling' (MBT) to distinguish the collective tunnelling of macroscopic objects from one-particle macroscopic quantum tunnelling.

In symmetric systems with more than one classical degree of freedom, tunnelling may be affected by the onset of dynamical instability and chaos. The mixed phase space on a given  energy shell typically contains two or more symmetry-related islands separated by a chaotic region. Motion between the different dynamical domains is classically forbidden, not by potential barriers but by residual motional constants. Dynamical quantum tunnelling \cite{Davis81,Keshavamurthy11} is allowed however, due to the weak quantum coupling across the chaos border. Chaos-supported states can then mediate between the regular islands and enhance or suppress tunnelling by orders of magnitude \cite{Tomsovic94,Leyvraz96,Dembowski00,Brodier01,Podolsky03,Eltschka05,Hofferbert05,Steck01,Hensinger01,Wuster12,Lenz13,Backer08,Doggen17,Arnal20,Martinez21}. Due to the resonant nature of this mechanism, the tunnelling rate becomes sensitive to the variation of control parameters or of the Planck constant, and exhibits large amplitude fluctuations with an overall Cauchy statistics \cite{Leyvraz96}. In the presence of Anderson localization within the chaotic sea, a transition is observed from Cauchy to log-normal distribution \cite{Doggen17}.

\sect{Outline}
In this work, we show how chaos may be employed to enhance many-body tunnelling. Considering a bi-partite cold atoms system, we explore its classical mixed phase space structure and its manifestations on the quantum many-body spectrum. We identify the chaos-island resonances, study the enhancement of many-body tunnelling, demonstrate quantum fluctuations in the dependence of the tunnelling gap on system parameters and particle number, and propose a dynamical scenario to verify the effect.

%%%%%%%%%%%%%%%%%%%%%%%%%
\sect{The double-dimer model}
Consider a system of two weakly-coupled bosonic Josephson junctions ('dimers'), described by the four-mode Hamiltonian,
\beq
\hat{H}&=& -\frac{\Omega}{2}\left( \sum_\alpha \hat{a}_{+,\alpha}^\dag\hat{a}_{-,\alpha} + H.c\right)+ \frac{U}{2}\sum_{\alpha,\sigma}\hat n_{\sigma,\alpha}\left(\hat{n} _{\sigma,\alpha}-1\right)\nonumber \\
&-& \frac{\kappa}{2} \left(\sum_\sigma \hat{a}^\dag_{\sigma,{\rm L}}\hat{a}_{\sigma,{\rm R}}+\hat{a}^\dag_{\sigma,{\rm R}}\hat{a}_{\sigma,{\rm L}}\right)
\label{Dicke_ham}
\eeq
where, $\hat a_{\sigma,\alpha}$ annihilate a boson in the $\sigma=\pm$ mode of the $\alpha={\rm L},{\rm R}$ junction. The intra-dimer (inter-dimer) hopping frequencies is  $\Omega$ ($\kappa$) and $U$ is the on-site interaction strength. Below we use dimensionless units in which $\Omega=1$ and $t\rightarrow \Omega t$ so that the dimensionless system parameters are $\omega=\kappa/\Omega$ and $u=UN/\Omega$. 

Accounting for $N$ conservation, the Hilbert space dimension of the many-body system is $D=(N+1)(N+2)(N+3)/6$. The classical mean-field dynamics has $d=3$ degrees of freedom, e.g. three population imbalances and three adjoint relative phases acting as adjoint action-angle variables. In the adiabatic limit $\omega\ll 1$, the timescale separation between fast intra-dimer motion and slow inter-dimer motion allows the elimination of internal degrees of freedom via a series of canonical transformations \cite{Strzys10,Strzys12,Khripkov13,Khripkov14,Ray20}. The dynamics is then reduced to the slow motion of the population imbalance $n = n_{\rm L}-n_{\rm R}$ where $n_{\alpha}=\sum_{\sigma} n_{\sigma,\alpha}$ and the action imbalance $j=j_{\rm L}-j_{\rm R}$ where $j_{\alpha}$ are single-dimer classical actions \cite{Sup}, and their adjoint phases. While the total particle number $N=n_{\rm L}+n_{\rm R}$ is strictly conserved, the total action $J=j_{\rm L}+j_{\rm R}$ is an adiabatic invariant that translates quantum mechanically to the total number of Josephson excitations. The two weakly-coupled subsystems thus exchange particles and excitations on timescales that are long compared to the inverse of the internal Josephson frequencies.

%%%%%%%%%%%%%%%%%%%%%%%
\begin{figure}
\centering
\includegraphics[clip=true, width=\columnwidth]{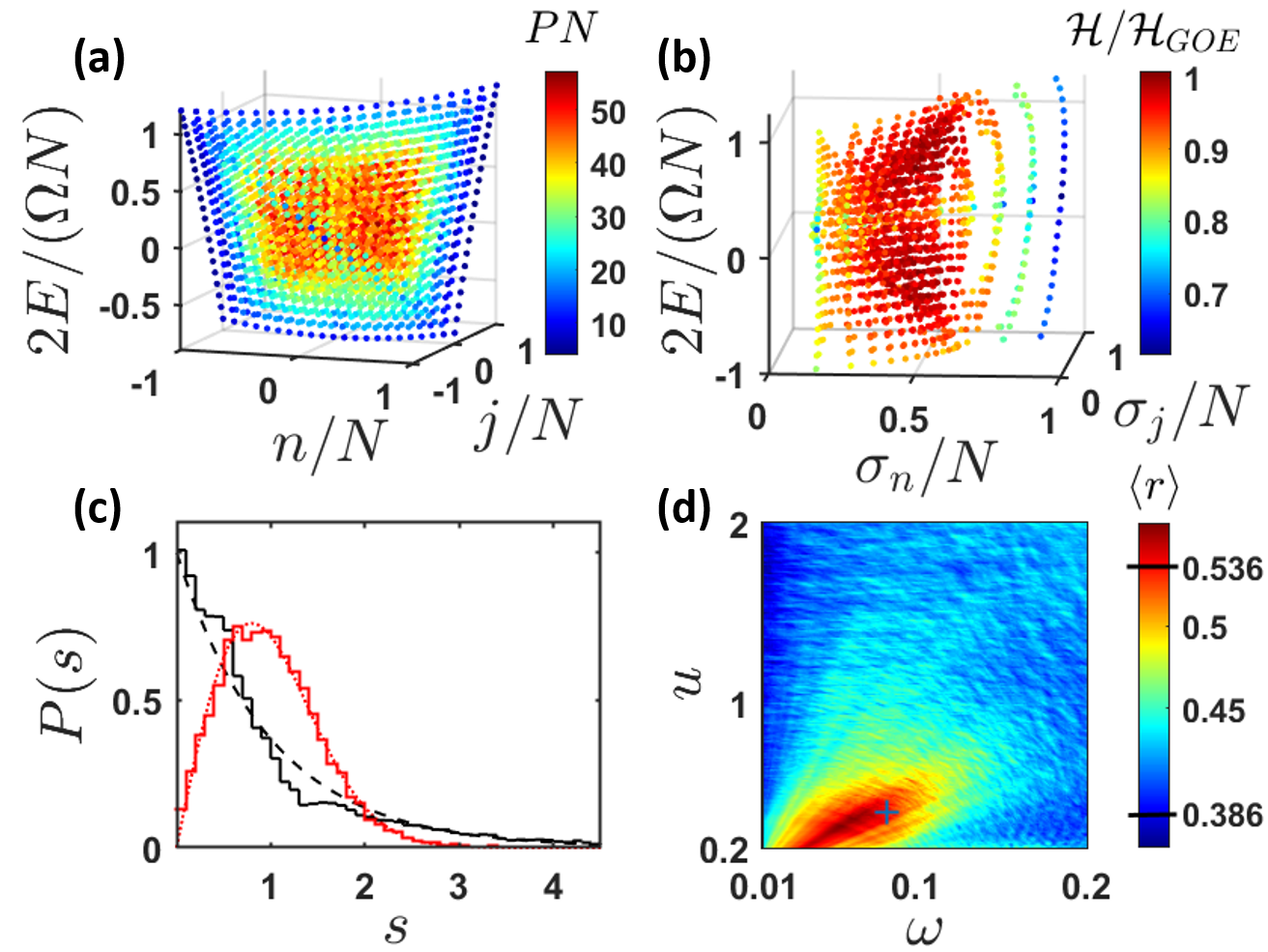}
\caption{{\bf Spectrum and level spacing statistics:} 
(a) The unperturbed ($\omega=0$) eigenenergies, classified according to the good quantum numbers $n,j$ and colored according to their participation number in the exact eigenstate basis for $\omega=0.082$; 
(b) The exact eigenenergies for the same $\omega$, classified according to the $\sigma_n,\sigma_j$ standard deviations and colored according to the Shannon entropy of the corresponding eigenstates; 
(c) Level spacing statistics for $\omega=0.01$ (black) and $\omega=0.082$ (red). For comparison, Poisson (dashed black) and Wigner surmise of GOE class (dotted red) are overlaid; 
(d) The spacing correlation measure $r$ throughout the parameter space. The expected values for a GOE and Poisson statistics are marked on the color bar. The $+$ marker marks the parameters used in panels (a)-(c). 
The interaction parameter for panels (a)-(c) is $u=0.4$. Here and in other figures we set particle number $N=21$ unless it is otherwise mentioned.}
\label{f1}
\end{figure}
%%%%%%%%%%%%%%%%%%%%%%%%%%%%%%

%%%%%%%%%%%%%%%%%%%%%%%
\begin{figure}
\centering
\includegraphics[clip=true, width=\columnwidth]{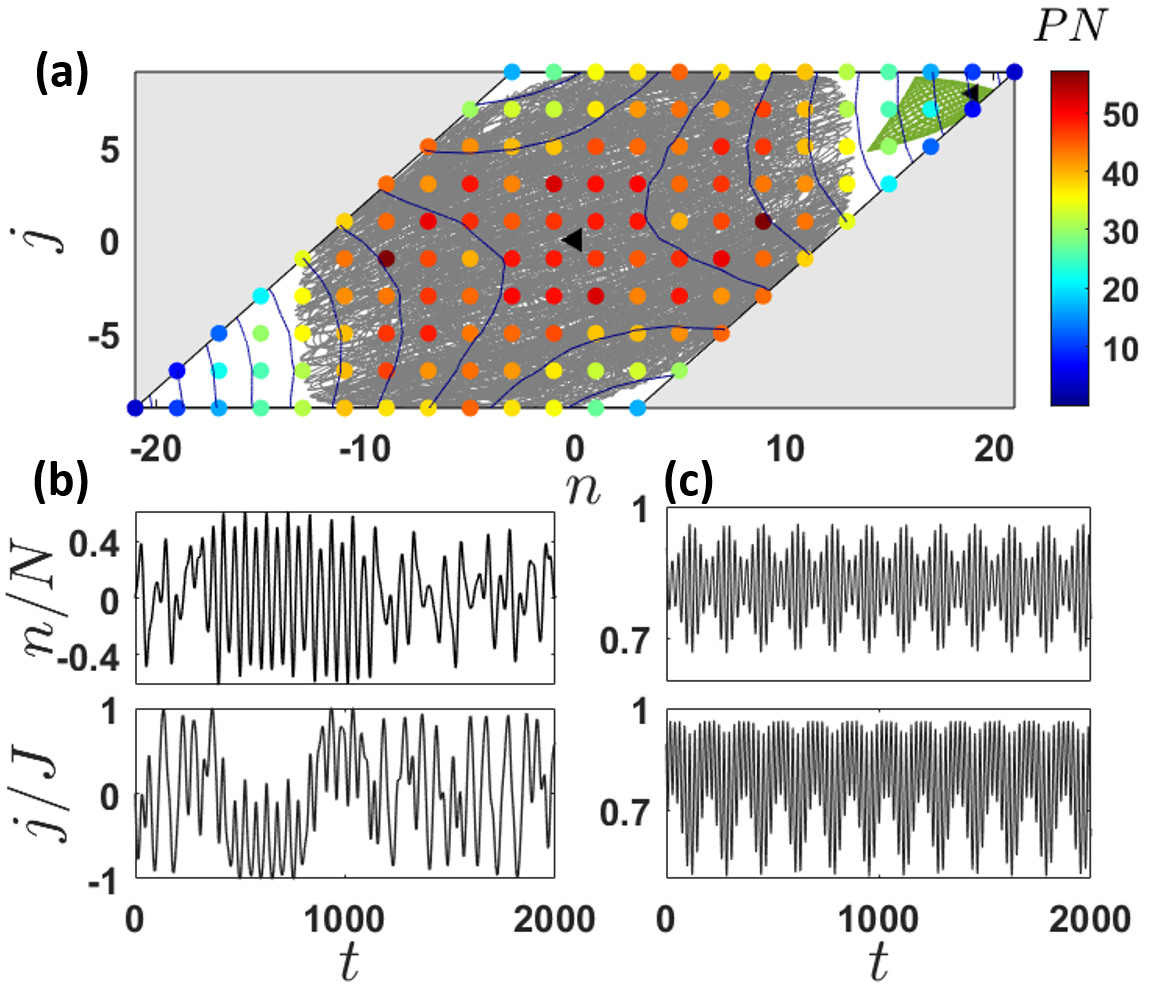}
\caption{{\bf Quantum-classical correspondence:} 
(a) The $J=9$ surface of the spectrum in Fig.~\ref{f1}a over representative chaotic (grey, center) and self-trapped (green, r.h.s.) classical trajectories launched at the points marked by filled triangles. Contour lines correspond to the energy $E(N,J,n,j)$. The adiabatic invariance of $J$ dissects the energy surface into self-trapped and chaotic regions. High quantum participation ratios correspond to chaotic classical dynamics since the exact chaotic eigenstates are spread throughout the entire chaotic region. The dynamics of the particle imbalance $n$ and action imbalance $j$ for the same chaotic and regular trajectories are plotted in (b) and (c), respectively.}
\label{f2}
\end{figure}
%%%%%%%%%%%%%%%%%%%%%%%%%%%%%%

\sect{Energy spectrum}
In Fig~\ref{f1}a we plot the energies of the $\omega=0$ unperturbed eigenstates $|\mu\rangle \equiv |N,J,n,j\rangle$, classified according to the good quantum numbers $n$ and $j$. The spectrum is arranged in layers corresponding to $J=0,\dots,N$. For example, the lowest (highest) energy states correspond to the direct product of ground state (fully excited) dimers with varying distributions of the $N$ particles between them. Within each fixed $J$ surface, states with high $|n|$ have higher energy due to the repulsive interaction between particles. By contrast, states with high $|j|$ have lower energies due to the effective attraction between excitations for $0<u<1$ \cite{Sup}.

The exact eigenenergies of the coupled dimers with $\omega=0.082$ are plotted in Fig.~\ref{f1}b. With finite coupling, the eigenstates belong to one of four irreducible representations ${\rm (A, B_1, B_2, B_3)}$ of the dihedral group $D_2$ depending on whether they are odd or even with respect to the symmetry operations $L\leftrightarrow R$ and $+\leftrightarrow -$. Hence $\langle n\rangle=\langle j\rangle=0$ and the states have to be arranged according to their standard deviations $\sigma_n=\sqrt{\langle n^2\rangle}$,  $\sigma_j=\sqrt{\langle j^2\rangle}$. The high $\sigma_n$ and high $\sigma_j$ states are nearly degenerate MSC doublets, consisting of even and odd superposition of the $\omega=0$ localized states, hence their $\sigma_{n,j}$  variance is close to the $|n|,|j|$ of the constituent unperturbed states. Considerable mixing is evident for states with lower $\sigma_{n,j}$.

%%%%%%%%%%%%%%%%%%%%%%%
\begin{figure}
\centering
\includegraphics[clip=true, width=\columnwidth]{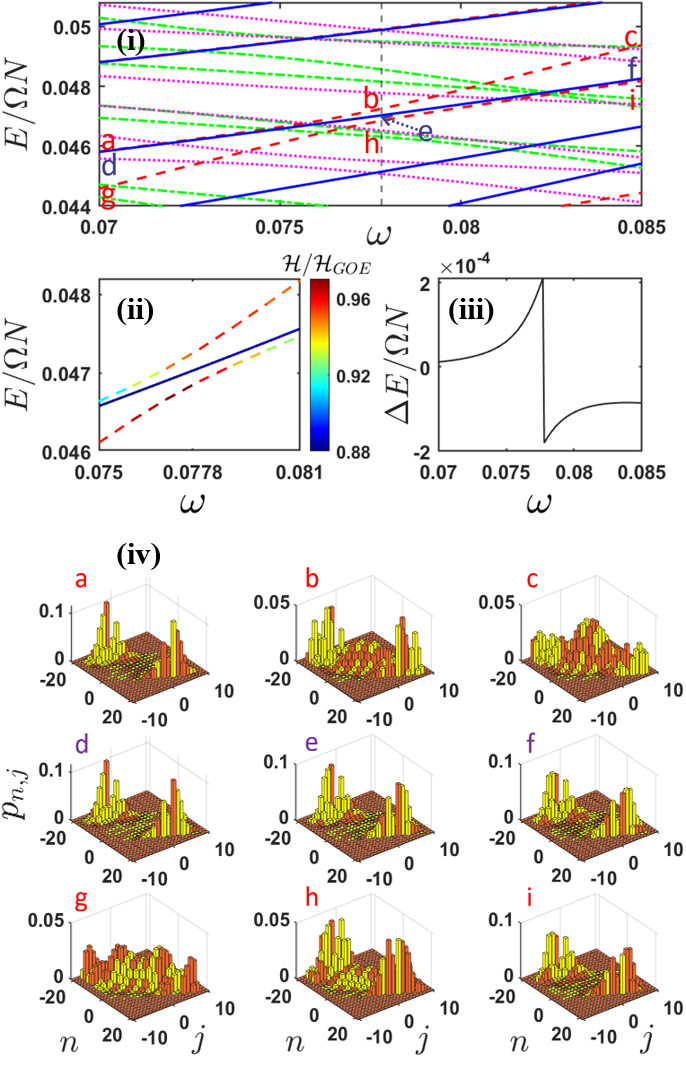}
\caption{{\bf Chaos assisted many-body tunnelling:} 
(i) Section of the eigenenergy spectrum with $u=0.4$ and $N=21$, plotted as a function of the coupling frequency $\omega$. Lines are colored according to the symmetry class of the corresponding eigenstates: A (dash-dotted green), ${\rm B_1}$  (dashed red), ${\rm B_2}$ (solid blue), and ${\rm B_3}$ (dotted magenta); 
(ii) Zoom on the $\{\rm B_1, B_2\}$ chaos-assisted MBT resonance with eigenenergy lines colored according to the Shannon entropy of the three participating eigenstates; 
(iii) The resulting chaos enhanced tunnelling gap between the doublet states; 
(iv)  The  $p_{n,j}$ distribution of the pertinent states, marked in (i) at $\omega=0.07$ (a,d,g), 0.0778 (b,e,h), and 0.085 (c,f,i). Bars are colored according to the sign of the (real) amplitudes.}
\label{f3}
\end{figure} 
%%%%%%%%%%%%%%%%%%%%%%%%%%%%%%

\sect{Onset of chaos}
Expanding the decoupled 2-dimer eigenstates $|\mu \rangle$ in the basis of exact finite-$\omega$ eigenvectors $|\nu\rangle$, we can calculate the participation number ${\rm PN}_\mu=1/(\sum_{\nu=1}^D p_{\mu,\nu}^2)$  where $p_{\mu,\nu}=|\langle\nu|\mu\rangle|^2$. The states in Fig~\ref{f1}a are colored according to their ${\rm PN}$ at $\omega=0.082$. While the high $n$ and high $j$ states at intermediate energy involve only a few (minimally two) exact eigenstates, there exists a large region of states with participation numbers that are comparable to the number of states in the fixed $J$ shell. This is typical to chaotic ergodization where the exact eigenstates are spread over an entire chaotic region within the pertinent energy shell. Chaos is also evident in Fig~\ref{f1}b where the exact eigenstates are colored according to their {\em Shannon entropy} ${\cal H}(\nu)=-\sum_{m=1}^D p_{\nu,m}\ln p_{\nu,m}$, where $p_{\nu,m}=|\langle m|\nu\rangle|^2$ are the probabilities in the computational Fock state basis $\{|m\rangle=|n_{L,+},n_{L,-},n_{R,+},n_{R,-}\rangle\}$. The maximal value of ${\cal H}=\ln D$ is obtained for a uniform distribution $p_{\nu,m}=1/D$. For a fully chaotic phase space, the eigenvector elements may be replaced by independent real random variables from a Gaussian distribution fluctuating around $1/D$, resulting in a lower Shannon entropy ${\cal H}_{\rm GOE}=\ln(0.48 D)$ \cite{Izrailev16}. By contrast, quasi-integrable localization gives much lower values of ${\cal H}$. It is thus clear from Fig~\ref{f1}b that the phase space of the double dimer is mixed, containing both chaotic and quasi-regular regions.

Level spacing statistics is presented in  Fig~\ref{f1}c (see also \cite{Sup}). In the extreme weak coupling limit $\omega\ll u$, the dynamics is dominated by self trapping and the spacing statistics is nearly Poissonian. At larger values of $\omega$, however, with $\omega \ll 1$, the spacing statistics changes to the Wigner surmise indicating the onset of chaos. Further increase in $\omega$ increases the linearity of the system so as to reinstate regular dynamics \cite{Sup}. To determine the parameter regime for observing chaos, we plot in Fig~\ref{f1}d the spacing correlation measure \cite{Bogomolny13}, defined as $r=\langle \min(s_n,s_{n+1})/\max(s_n,s_{n+1}\rangle$ where $\{s_n\}$ are adjacent level spacing and $\langle .\rangle$ denotes averaging over all spacing, as a function of $\omega$ and $u$. The value $r\approx0.536$ indicates the GOE random-matrix statistics anticipated for time-reversal-invariant systems exhibiting chaotic dynamics, whereas, the value $r\approx0.386$ indicates the Poissonian statistics of systems undergoing regular dynamics. 

%%%%%%%%%%%%%%%%%%%%%%%
\begin{figure}
\centering
\includegraphics[clip=true, width=\columnwidth]{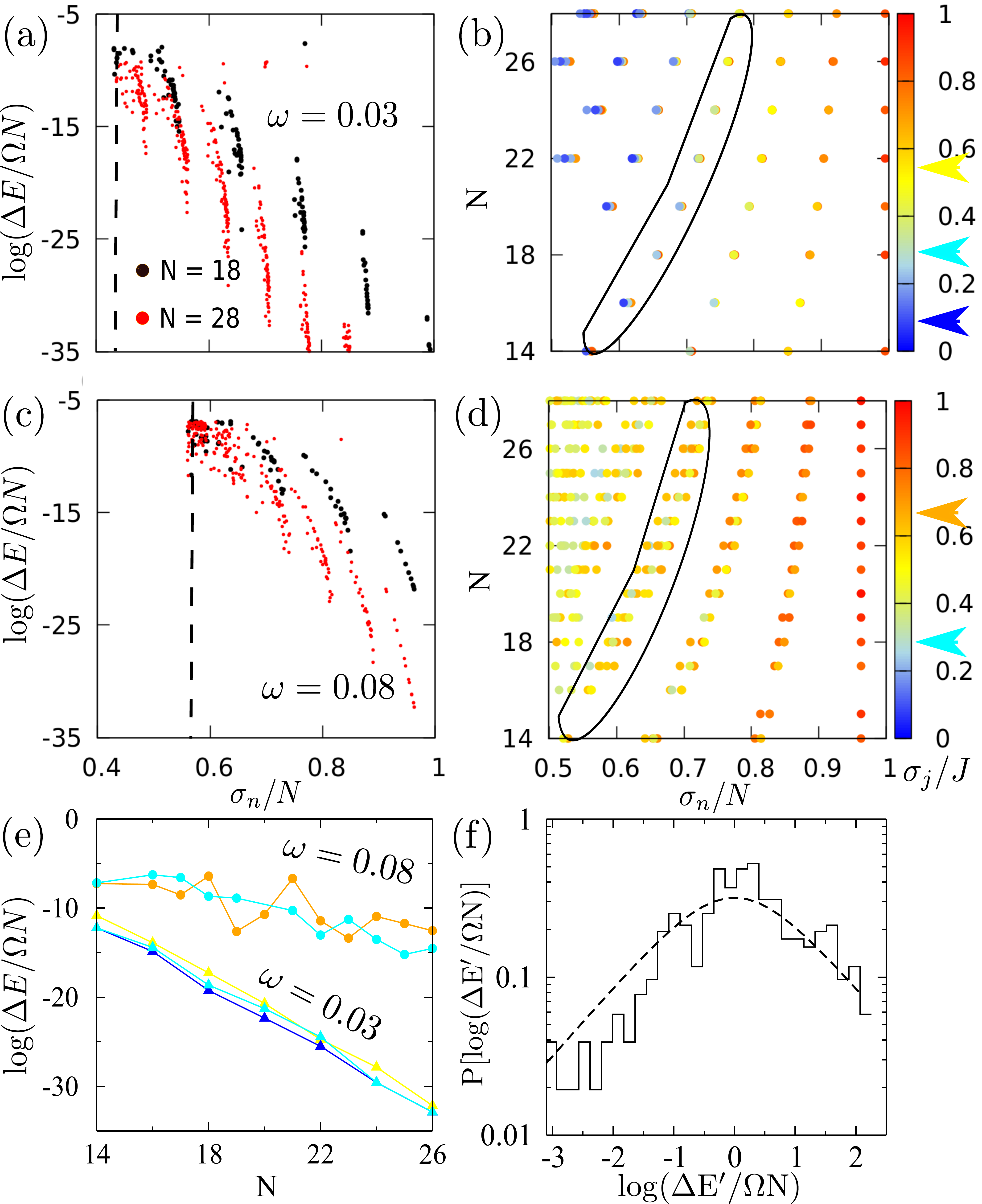}
\caption{{\bf Tunnelling rate fluctuations.} 
(a) Tunnel gaps $\Delta E/\Omega N$ between doublet states, belonging to $\{\rm B_1, B_2\}$ symmetry classes, are plotted against $\sigma_n/N$ for $\omega=0.03$. Dashed line marks the island-chaos border (b) The points represent one of the states in $\{\rm B_1, B_2\}$ doublet lying on the $J/N \approx 1/2$ surface and having the larger $\sigma_n/N$ of the two, arranged according to $\sigma_n/N$ with varying $N$, and colored according to their $\sigma_j/J$; 
(c-d) same as in (a-b) for $\omega=0.08$; 
(e) $N$-dependence of the doublet spacing. Point color corresponds to the color code in panels (b) and (d) at the marked places. Triangles and circles correspond to the encircled group in panel (b) and (d), exhibiting smooth exponential decay and fluctuations, respectively.
(f) Statistics of fluctuations $\Delta E'$ around the mean spacing $\overline{\Delta E}$ corresponding to doublets encircled in (d) for each $N$, varied up to $N=46$.
}
\label{f4}
\end{figure} 
%%%%%%%%%%%%%%%%%%%%%%%%%%%%%%

\sect{Self-trapping of particles and excitations}
One fixed $J$ surface of the spectrum in Fig.~\ref{f1}a and representative classical trajectories (Poincar\'e sections are not possible due to the 4D motion within the fixed $\{E,J\}$ shell) at the pertinent energies are overlaid in Fig.~\ref{f2}. The mutual conservation of the total energy $E$ and the total action $J$ dissects the classical energy shell into symmetry related pairs of regular islands where either particles or excitations are self-trapped by their interactions. The islands are separated by a chaotic region, ergodically explored by classical trajectories dominated by the saddle point at $n=j=0$. The high participation quantum states are supported by the chaotic region, whereas the low participation states are supported by the islands. This is the standard configuration for chaos assisted tunnelling.

\sect{Chaos assisted many-body tunnelling}
In Fig.~\ref{f3}(i) we plot a small section of the many-body spectrum vs the coupling parameter $\omega$ in the chaos regime, with eigenstates classified according to their symmetry. Level repulsion within each symmetry class is evident, whereas, states with different symmetries freely cross one another. The zoom-in in Fig.~\ref{f3}(ii) demonstrates the standard scenario of chaos assisted tunnelling, wherein a $\{{\rm B_1},{\rm B_2}\}$ doublet consisting of low-$\cal H$ MSC-like states encounters a high-$\cal H$ chaos-supported ${\rm B_1}$ state. The state belonging to the ${\rm B_1}$ class is repelled, whereas, the ${\rm B_2}$ state remains unaffected, resulting in substantial enhancement of the doublet tunnelling gap, see Fig.~\ref{f3}(iii). The probability distributions $p_{n,j}=\sum_J|\langle N,J,n,j|\psi\rangle|^2$ for the pertinent states at the positions marked in panel (i), are shown in Fig.~\ref{f3}(iv)a-i. The identity of the the ${\rm B_1}$ MSC state changes through resonance, resulting in the sign change in panel (iii), and substantial mixing between the island and chaos states is observed on resonance. 

%%%%%%%%%%%%%%%%%%%%%%%
\begin{figure}
\centering
\includegraphics[clip=true, width=\columnwidth]{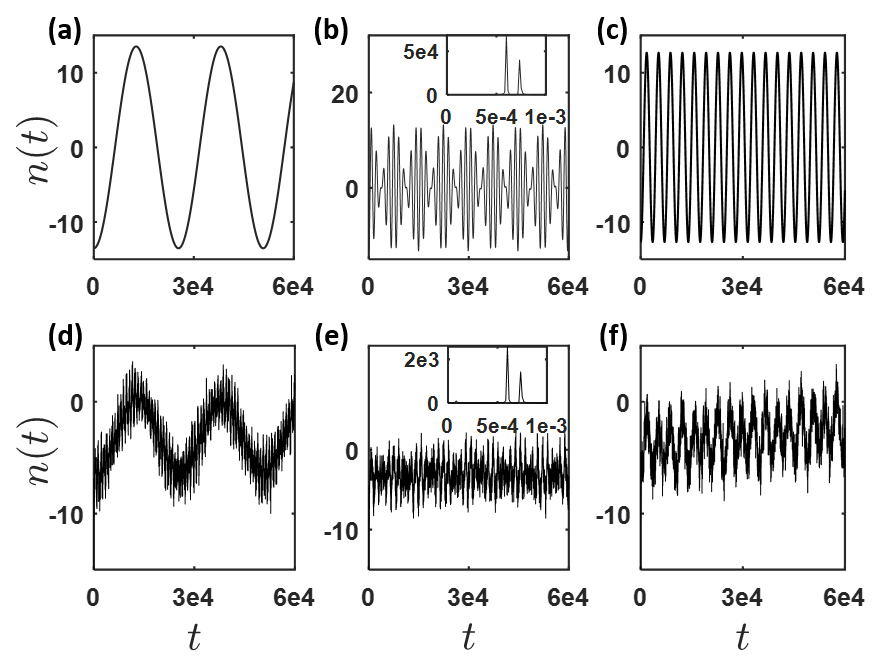}
\caption{{\bf Signature of chaos assisted tunneling in the population imbalance dynamics:}
Panels (a)-(c) depict the tunneling dynamics for $\omega=0.07$ (a), 0.0778 (b), and 0.85 (c) for an ideal local preparation using only the states involved in the resonance shown in Fig.~\ref{f3}. 
Panels (d)-(f) show a more realistic scenario where the system is prepared at the unperturbed eigenstate $|N,J,n,j\rangle=|21,9,-15,-3\rangle$ (in d,e) or $|21,9,-15,-5\rangle$ (f) and $\omega$ is then quenched to the stated value. 
The insets in (b),(e) show the dual frequency power spectra on resonance.}
\label{f5}
\end{figure}
%%%%%%%%%%%%%%%%%%%%%%%%%%%%%%

\sect{Fluctuations} The dependence of doublet spacing on the total particle number $N$ is presented in Fig.~\ref{f4}. The regular island doublets for a given $N$ are arranged in groups according to $\sigma_n$ and the group mean tunneling gap decays exponentially with $\sigma_n$ [cf. Fig.~\ref{f4}(a,c)]. Plotting $\sigma_n$ of doublets on a fixed $J=N/2$ surface with varying $N$, they group according to $\sigma_n/N \approx 1-i/N$, where $i=0,1,...i_0$ and $i_0$ is the group closest to the chaos border, as displayed in Fig.~\ref{f4}(b,d). Focusing on a single group, the tunnelling rate of doublets with the same $\sigma_j/J$ decays exponentially with the effective inverse Planck constant $N$ in the absence of chaos. By contrast, when the doublet states are coupled to the chaotic manifold, the tunnelling rate fluctuates erratically over several orders of magnitude about the exponentially decreasing mean doublet spacing $\overline{\Delta E}$, see Fig.~\ref{f4}e. The resulting statistics of doublet spacing fluctuations, namely, $\Delta E'=\Delta E-\overline{\Delta E}$ is compared in Fig.~\ref{f4}f with the anticipated Cauchy distribution \cite{Doggen17}, see \cite{Sup} for details.

\sect{Dynamical signature of chaos assisted many-body tunneling}
The three-level nature of chaos assisted tunnelling is evident in the population-imbalance dynamics presented in Fig.~\ref{f5}. Away from resonance, a localized preparation consists only the two doublet states, hence  the tunneling oscillations involve a single frequency determined by the doublet spacing (Fig.~\ref{f5}a,c). At the resonance, the mixing of one of the doublet states with the same-symmetry chaos-supported state implies all three states are required for a localized preparation, hence the tunneling oscillations exhibit two-frequency beating (Fig.~\ref{f5}b). This signature of chaos assisted tunneling in the population dynamics is retained in a more realistic scenario where the two dimers are first prepared separately (i.e. the initial state is an eigenstate of the unperturbed system) and the coupling frequency $\omega$ is then quenched to the pertinent value (see Fig.~\ref{f5}d-f). While the latter preparation  projects onto other eigenstates supported by the integrable island region, the population dynamics is dominated by the resonant states.

\sect{Conclusions}
We conclude that chaos may assist in the observation of tunneling on macroscopic scales. The main impediment towards the realization of many-body tunneling  is the poor scaling of the $N$-particle tunneling rate with the number of particles. The huge enhancement of tunneling by chaos-induced resonances, as translated to the many-body bosonic system, may greatly increase the tunneling object size and push the limits of macroscopic quantum tunneling. 

\sect{Acknowledgment}
This research was supported by the Israel Science Foundation (Grant No.283/18). SR acknowledges a scholarship from the Alexander von Humboldt Foundation, Germany.

%%%%%%%%%%%%%%%%%%%%%%%%%%%%%%%%%%%%%%%%%%%%%%%%%%%%%%%%%%%%%%%%%%%%%%%%%%%%%%%%%%%%%%%%%%%%%%%%%%%%%%%%%%%
%%%%%%%%%%%%%%%%%%%%%%%%%%%%%%%%%%%%%%%%%%%%%%%%%%%%%%%%%%%%%%%%%%%%%%%%%%%%%%%%%%%%%%%%%%%%%%%%%%%%%%%%%%%
\clearpage
\onecolumngrid
\pagestyle{empty}

\renewcommand{\thefigure}{S\arabic{figure}}
\setcounter{figure}{0}
\renewcommand{\theequation}{S-\arabic{equation}}
\setcounter{equation}{0}

\Cn{{\large \bf Chaos assisted many-body tunneling}} 

\Cn{{Urbashi Satpathi, Sayak Ray, and Amichay Vardi}} 

\Cn{{\large (Supplementary Material)}} 

Below we review the derivation of analytical expressions for the classical action of a Josephson junction and the effective interaction between the Josephson excitations. We further discuss the details of the procedures adapted while computing spectral statistics for the detection of chaos and the statistics of doublet-spacing fluctuations which is the signature of chaos assisted tunnelling.

\rmrk{\section{Classical action of a Bose-Hubbard dimer for $0\leq u<1$}}
\label{josons}

We consider the single-dimer Hamiltonian, $\hat{H}_\alpha$ , i.e. the ${\alpha=\rm L}$ or ${\rm R}$  terms in the Hamiltonian~(\ref{Dicke_ham}) with $\kappa=0$, conserving the particle number $n_\alpha$. Using the Schwinger representation of angular momentum with the total angular momentum quantum number $L_\alpha=n_\alpha\hbar/2$, $\hat{H}_\alpha$ can be re-written as (omitting the subscript $\alpha$ for simplicity),
\begin{equation}
\frac{\hat{H}}{\Omega} = -\hat{L}_x + \frac{u}{2L}\hat{L}_x^2 + \frac{uL}{2}
\end{equation}     
where, $\hat{L}_x=(\hat{a}_{+}^{\dagger}\hat{a}_{-}+\hat{a}_{-}^{\dagger}\hat{a}_{+})/2$, $\hat{L}_y=(\hat{a}_{+}^{\dagger}\hat{a}_{-}-\hat{a}_{-}^{\dagger}\hat{a}_{+})/(2i)$ and $\hat{L}_z=(\hat{n}_{+}-\hat{n}_{-})/2$ are the $\rm SU(2)$ generators. Promoting the operators to the classical variables, and considering the fact that the last term $uL/2$ is a constant of $\hat{H}$, the classical energy can thus be written as,
\begin{equation}
H(P,\phi) = -P + \frac{u}{2} \frac{L^2-P^2}{L}\sin^2 \phi
\end{equation}
where, we have introduced the action-angle variable as follows: $L_x=P$, $L_y=\sqrt{L^2-P^2}\cos \phi$, $L_z=\sqrt{L^2-P^2}\sin \phi$, for $-L \le P \le L$. Thus the phase space area enclosed by an orbit of energy $H(P,\phi)=E$ is given by,
\begin{equation}
A(E) = 2\pi L - \oint d\phi P(E,\phi)
\end{equation}
where, solving the quadratic equation, and discarding one branch because we must have $P>-L$, gives
\begin{equation}
P(E,\phi) = \frac{L}{u\sin^2\phi}\left(-1 + \sqrt{1 - 2u\frac{E}{\Omega L}\sin^2\phi + u^2\sin^4\phi}\right)
\end{equation}
Within the Bohr-Sommerfeld quantization, the total number of Josephson excitation for a given energy $E$ is $M(E)=A(E)/(2\pi\hbar)$. Since for $u<1$, the energy $E$ ranges between $\mp L\Omega$, therefore, we can parametrize $E=\Omega L \sin \eta$. Thus we can write $P(E,\phi)$ in terms of $\eta$ and $\phi$ as follows,
\begin{equation}
X(\eta,\phi) = \frac{P(\eta,\phi)}{L} = \frac{1}{u\sin^2\phi}\left(-1 + \sqrt{(1-u\sin \eta \sin^2 \phi)^2 + u^2\sin^4\phi}\right)
\end{equation}
We can now re-write, $M(n,\eta)=n j(\eta)$, where $j(\eta)=\frac{1}{2}-\frac{1}{4\pi}\oint d\phi X(\eta,\phi)$. Defining the new integration variable, $z=\sin^2\phi$ and then integrating by parts allows us to write $j(\eta)$ as
\begin{equation}
j(\eta) = \frac{1}{2} - \frac{1}{\pi} \int^{1}_{0} dz \sqrt{\frac{1-z}{z}} \frac{uz-\sin \eta}{\sqrt{(1+iue^{i\eta}z)(1-iue^{i\eta}z)}} 
\end{equation} 
This integral can be reduced further in terms of complete Elliptic integrals as discussed in detail in Ref.~\cite{Ray20} and the final expression reads,
\begin{align}
j =&\frac{1}{2} - \frac{1}{\pi}\mathrm{Re}\left[ 2\cos\eta\frac{K\!\!\left(\frac{1-i e^{-i \eta } u}{1+ie^{i \eta }u}\right)}{\sqrt{1+i e^{i \eta } u}}+\frac{2i}{u}\sqrt{1+i e^{i \eta } u}\, E\!\!\left(\frac{1-i e^{-i \eta} u}{1+ie^{i \eta }u}\right)
 +\frac{i e^{i \eta } \left(1-i e^{-i \eta } u\right) \Pi\!\!\left(i e^{-i \eta } u|\frac{2 i u \cos  \eta }{1+ie^{i \eta } u}\right)}{\sqrt{1+i e^{i \eta } u}}\right],
 \label{jdef}
\end{align}
where $K(m)$, $E(m)$, and $\Pi(n|m)$ are respectively the complete elliptic integrals of the 1st, 2nd and 3rd kinds. To avoid notational confusion, note that '$j$' in Eq.~(\ref{jdef}) refers to either of the one dimer actions $j_\alpha(n_\alpha,E_\alpha)$ rather than to their difference  $j_{\rm L} (n_{\rm L},E_{\rm L})- j_{\rm R} (n_{\rm R},E_{\rm R})$ as in the main text. 

\rmrk{\section{Josephson oscillations of particles and excitations}}

In this section, we briefly review the derivation by Strzys and Anglin in Ref.~\cite{Strzys10}, reducing the adiabatic four-mode dynamics into the slow Josephson oscillations of particles and excitations (`josons') between the two bosonic Josephson junctions. We begin by stating the following Holstein-Primakoff transformation (HPT) that we are going to apply in Eq.~\ref{Dicke_ham}.
\begin{subequations}
\begin{eqnarray}
\frac{n_{\alpha}}{2}-\hat{\mathcal{A}}_\alpha ^{\dagger} \hat{\mathcal{A}}_\alpha &\equiv & \frac{1}{2}(\hat{a}_{\alpha,+}^{\dagger}\hat{a}_{\alpha,-}+\hat{a}_{\alpha,-}^{\dagger}\hat{a}_{\alpha,+}) \\
\sqrt{n_{\alpha}-\hat{\mathcal{A}}_\alpha ^{\dagger} \hat{\mathcal{A}}_\alpha} \hat{\mathcal{A}}_\alpha &\equiv & \frac{1}{2}(\hat{a}_{\alpha,+}^{\dagger}+\hat{a}_{\alpha,-}^{\dagger})(\hat{a}_{\alpha,+}-\hat{a}_{\alpha,-}) \\
\hat{\mathcal{A}}_\alpha^{\dagger} \sqrt{n_{\alpha}-\hat{\mathcal{A}}_\alpha ^{\dagger} \hat{\mathcal{A}}_\alpha} &\equiv & \frac{1}{2}(\hat{a}_{\alpha,+}^{\dagger}-\hat{a}_{\alpha,-}^{\dagger})(\hat{a}_{\alpha,+}+\hat{a}_{\alpha,-}) 
\end{eqnarray}  
\label{HPT}
\end{subequations}
where, $\hat{\mathcal{A}}_\alpha$ describes the moving of atoms between the two $\sigma=\pm$ modes of a dimer labelled by $\alpha=\rm \{L,R\}$, and it obeys the commutation relation, $[\hat{\mathcal{A}}_\alpha, \hat{\mathcal{A}}_\alpha^{\dagger}]=1$ and $[\hat{\mathcal{A}}_\alpha, \hat{n}_\alpha]=0$. Applying Eq.~\ref{HPT} to Eq.~\ref{Dicke_ham} followed by the standard Bogoliubov transformation, $\hat{\mathcal{A}}_\alpha=u_\alpha \hat{\mathcal{J}}_\alpha+v_\alpha \hat{\mathcal{J}}_\alpha^{\dagger}$, for diagonalizing the quadratic Hamiltonian, the single dimer Hamiltonian becomes,
\begin{eqnarray}
\hat{H}_{\alpha} &=& -\frac{\Omega}{2} (\hat{a}_{+,\alpha}^\dag\hat{a}_{-,\alpha} + {\rm h.c.})+ \frac{U}{2}\sum_{\sigma}\hat n_{\sigma,\alpha}\left(\hat{n} _{\sigma,\alpha}-1\right) \nonumber \\
&\rightarrow & \frac{\Omega}{2}\hat{n}_\alpha + \frac{U}{4}\hat{n}_\alpha (\hat{n}_\alpha-2)+ \sqrt{\Omega(\Omega +U\hat{n}_\alpha)}\hat{\mathcal{J}}_\alpha^{\dagger}\hat{\mathcal{J}}_\alpha - \frac{U}{8}\frac{4\Omega+U\hat{n}_\alpha}{\Omega+U\hat{n}_\alpha} \hat{\mathcal{J}}_\alpha^{\dagger 2}\hat{\mathcal{J}}_\alpha^2 + \mathcal{O}(Un_{\alpha}^{-1})
\label{HPT_ham}
\end{eqnarray}
where, $u_\alpha$ and $v_\alpha$ are quasi-hole and particle excitation amplitudes, respectively, while, $\hat{\mathcal{J}}_\alpha$ obeys the commutation relation, $[\hat{\mathcal{J}}_\alpha, \hat{\mathcal{J}}_\alpha^{\dagger}]=1$. While deriving Eq.~\ref{HPT_ham} we have also dropped out the terms which do not commute with $\hat{\mathcal{J}}_\alpha^{\dagger}\hat{\mathcal{J}}_\alpha$. Next, to take into account the weak coupling which is responsible for particle transfer between the two dimers, we apply the similar HPT, which in the large-N limit can be written as,
\begin{equation}
\hat{n}_{\rm L,R}=\frac{1}{2}[N\pm N^{1/2}(\hat{\mathcal{A}}^{\dagger}+\hat{\mathcal{A}})]
\label{HPT1}
\end{equation}  
where, $\hat{\mathcal{A}}$ describes the transfer of atoms between $\rm L$ and $\rm R$ junctions, and obeys the commutation relation, $[\hat{\mathcal{A}},\hat{\mathcal{A}}^{\dagger}]=1$. It can be noted that Eq.~\ref{HPT1} preserves the total number $n_{\rm L}+n_{\rm R}$. Therefore, the total Hamiltonian in Eq.~\ref{Dicke_ham}, including the single dimer $\hat{H}_\alpha$ and the coupling between them, can be written as (rescaled by $\Omega$),
\begin{equation}
\hat{H} \rightarrow \omega \hat{\mathcal{A}}^{\dagger}\hat{\mathcal{A}} + 
\frac{u}{8}(\hat{\mathcal{A}}^{\dagger}+\hat{\mathcal{A}})^2 - 
\frac{\omega_{\mathcal{J}}}{2}(\hat{\mathcal{J}}_{\rm L}^{\dagger}\hat{\mathcal{J}}_{\rm R} + \hat{\mathcal{J}}_{\rm R}^{\dagger}\hat{\mathcal{J}}_{\rm L}) - 
\frac{U_{\mathcal{J}}}{2}\sum_{\alpha=\rm L,R}\hat{\mathcal{J}}_{\alpha}^{\dagger 2}\hat{\mathcal{J}}_{\alpha}^2 + 
\frac{u}{4}\sqrt{\frac{1}{1+u/2}}\frac{(\hat{\mathcal{A}}+\hat{\mathcal{A}}^{\dagger})}{\sqrt{N}}(\hat{\mathcal{J}}_{\rm L}^{\dagger}\hat{\mathcal{J}}_{\rm L}-\hat{\mathcal{J}}_{\rm R}^{\dagger}\hat{\mathcal{J}}_{\rm R})
\label{joson_ham}
\end{equation}
where, the effective tunnelling frequency and interaction strength of the Josephson excitations are given by,
\begin{equation}
\omega_{\mathcal{J}} = \omega \frac{1 + u/4}{\sqrt{1+u/2}} \quad \text{and} \quad U_{\mathcal{J}} = U\frac{1+u/8}{1+u/2},
\end{equation}
respectively. The two first terms on the r.h.s. of Eq.~(\ref{joson_ham}) correspond to Josephson oscillations of particles whereas the third and fourth terms are a Josephson Hamiltonian for the excitations with effective attractive interaction between them. The last term couples the two oscillations (due to the dependence of the fast internal dimer frequencies on particle number). In addition to $N$, the total number of excitations $J=\sum_{\alpha={\rm L,R} }{\mathcal J}_\alpha^\dag {\mathcal J}_\alpha\rightarrow j_L+j_R$ is also conserved by the approximate Hamiltonian~(\ref{joson_ham}). While the conservation of $N$ is strict, $J$ is an adiabatic invariant.

%%%%%%%%%%%%%%%%%%%%%%%%%%%%%%%%%%%%%%%%%%%%%%%%%%%%%%%%%%%%%

\rmrk{\section{Level spacing statistics - spectrum unfolding procedure}}
\label{unfolding}

\begin{figure}
\centering
\includegraphics[clip=true, width=\columnwidth]{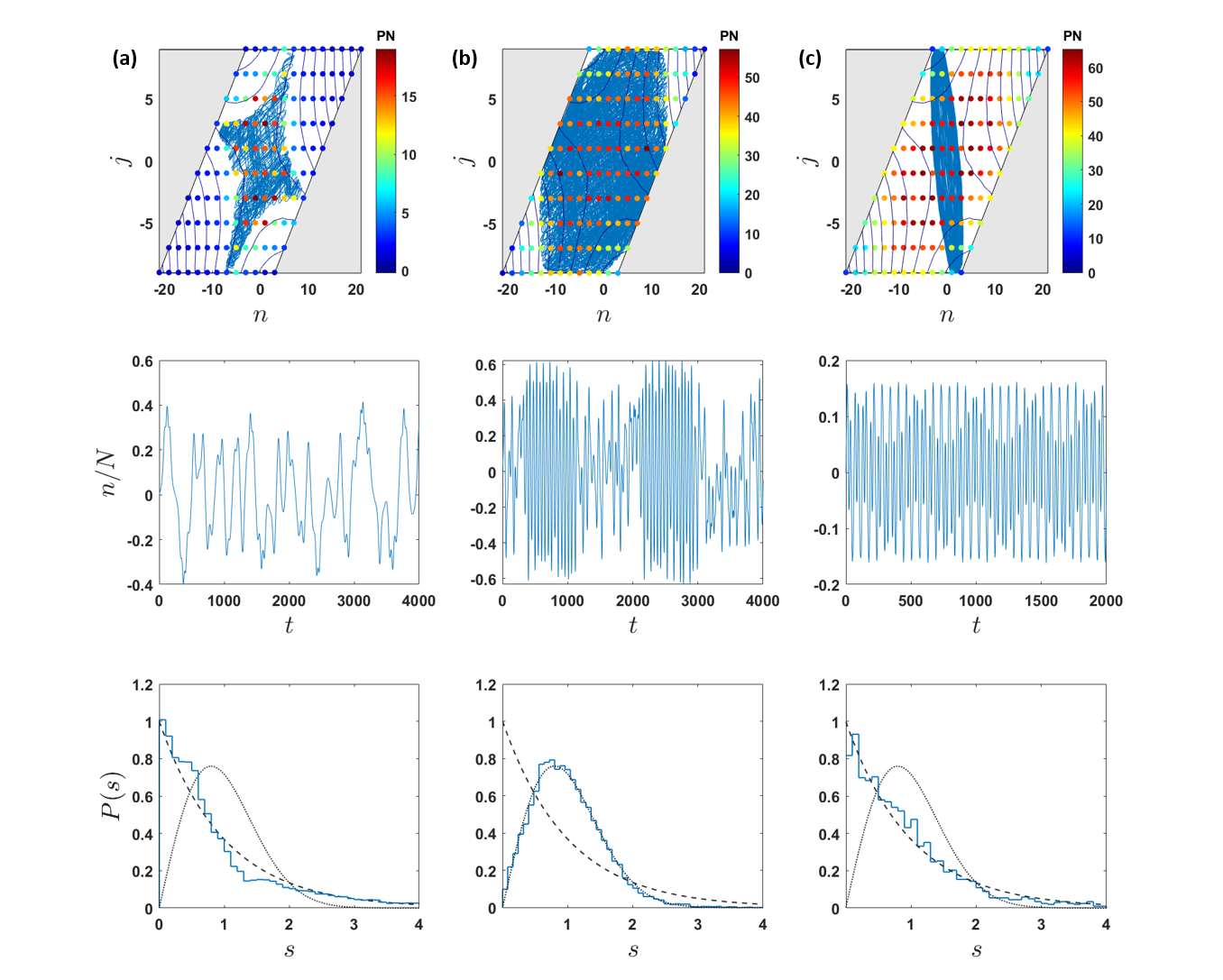}
\caption{Comparison of classical dynamics, quantum participation numbers, and level spacing statistics for: (a) $\omega=0.01$, (b) $\omega=0.082$, (c) $\omega=0.2$. The interaction strength is $u=0.4$. Top and middle panels as in Fig.~\ref{f2}, bottom panels as in Fig.~\ref{f1}c. The classical trajectory in (a) and (b)  is chaotic and conserves the total action $J$. By contrast, the trajectory in (c), launched at the same saddle point, is regular and does not remain on the fixed $J$ surface, indicating the breakdown of timescale separation.}
\label{fsm}
\end{figure}

%%%%%%%%%%%%%%%
The level spacing statistics is obtained by separating the spectrum to the four symmetry classes of the dihedral group $D_2$ and unfolding the resulting spectra according to the energy-dependent mean spacing $\bar s(E)$. This procedure eliminates artefacts of the non-uniform density of states which may be misinterpreted as level spacing correlations. 
Accordingly, we introduce a moving energy interval $\delta E$ which is large with respect to the mean spacing, i.e. $\delta E\gg \bar s(E)$ but small with respect to the energy scale on which the density of states varies, i.e. $\partial_E {\bar s}(E) \delta E\ll {\bar s}(E)$. Level spacing at energy $E$ are then expressed in units of the mean spacing within the interval $[E-\delta E/2,E+\delta E/2]$ and the four symmetry classes are binned together to give the histograms of Fig.~\ref{f1}. The agreement between the classical dynamics and the level spacing statistics is also demonstrated in Fig.~\ref{fsm}. For $\omega\ll u$ (Fig.~\ref{fsm}a) the dynamics is dominated by the regular self-trapping islands separated by a narrow chaotic sea. The resulting spacing statistics is nearly Poissonian $P(s)\approx e^{-s}$. The expansion of the chaotic sea with $\omega$ is reflected by a transition to the GOE Wigner-Dyson statistics $P(s)=(\pi/2) s e^{-\pi s^2/ 4}$ (Fig.~\ref{fsm}b). For 
larger values of $\omega$ the dynamics becomes quasi-linear, hence integrable, and Possonian spacing statistics is restored (Fig.~\ref{fsm}c).
%%%%%%%%%%%%%%%%%%%
 
 %%%%%%%%%%%%%%%%%%%%%%%
%%%%%%%%%%%%%%%%%%%%%%%%%%%%%%
 
\rmrk{\section{Tunneling rate fluctuation statistics}}
 \label{Cauchy}

%%%%%%%%%%%%%%%%%%%%%%%
In order to obtain the statistics of fluctuations in tunnelling rate with varying total particle number $N$ (equivalent to $1/\hbar$), we first compute the energy spacing $\Delta E$ between the doublet states belonging to $\{{\rm B_1},{\rm B_2}\} $ symmetry class. Note that the tunnelling rate between the doublet states is controlled by the effective inter-dimer coupling $\omega/u$ as demonstrated in Fig.~\ref{f3}. We filter out one of the near-chaos-border group of doublets lying on the $J/N \approx 1/2$ surface and arranged according to $\sigma_n/N$ as encircled in Fig.~\ref{f4}(b,d), and for each value of $N$ we subtract the mean spacing $\overline{\Delta E}$ of these group of doublets, which yields the fluctuations $\Delta E'=\Delta E-\overline{\Delta E}$. We have checked that the variance of fluctuations in $\Delta E'$ remains similar for all $N$. Therefore, without any loss of generality, we collect $\Delta E'$s for several values of $N$ and the resulting probability distribution of $\Delta E'$ is shown in Fig.~\ref{f4}f.         
%%%%%%%%%%%%%%%%%%%%%%%

%%%%%%%%%%%%%%%%%%%%%%%%%%%%%%%%%%%%%%%%%%%%%%%%%%%%%%%%%%%%%%%%%%%%%%%
\clearpage
\end{document}